\documentclass[preprint,12pt]{elsarticle}
\usepackage{amssymb,graphicx}

\def\x{\mathbf{x}}

\def\l{{\cal L}}

\def\pd{\partial}

\def\exp{\mathrm{exp}}
\def\ex{\mathrm{e}}
\def\be{\begin{equation}}
\def\ee{\end{equation}}
\def\bea{\begin{eqnarray}}
\def\eea{\end{eqnarray}}
\def\ie{\textit{i.e.} }

\journal{Communications in Theoretical Physics}

\begin{document}

\begin{frontmatter}

\title{The statistical properties of protein folding in the $\phi^4$ theory}

\author[ui]{M. Januar} 
\author[bppt]{A. Sulaiman\footnote{Email : albertus.sulaiman@bppt.go.id}} 
\author[ui,gftk]{L.T. Handoko\footnote{Email : handoko@teori.fisika.lipi.go.id,
laksana.tri.handoko@lipi.go.id}}
\address[ui]{Department of Physics, University of Indonesia,
Kampus UI Depok, Depok 16424, Indonesia\\}
\address[bppt]{Badan Pengkajian dan Penerapan Teknologi, BPPT Bld. II (19$^{\rm
th}$ floor), Jl. M.H. Thamrin 8, Jakarta 10340, Indonesia}
\address[gftk]{Group for Theoretical and Computational Physics,
Research Center for Physics, Indonesian Institute of Sciences,
Kompleks Puspiptek Serpong, Tangerang 15310, Indonesia}

\begin{abstract}
The statistical properties of protein folding within the $\phi^4$ model are
investigated. The calculation is performed using statistical mechanics and path 
integral method. In particular, the evolution of heat capacity in term of
temperature is given for various levels of the nonlinearity of source and the
strength of interaction between protein backbone and nonlinear source. It is
found that the nonlinear source contributes constructively to the specific heat
especially at higher temperature when it is weakly interacting with the protein
backbone. This indicates increasing energy absorption as the intensity of
nonlinear sources are getting greater. The simulation of protein folding
dynamics within the model is also refined.
\end{abstract}


\end{frontmatter}

\section{Introduction}

It is well known that the time ordered of protein folding is realized from the primary to the secondary and subsequent structures. Furthermore, the secondary structure consists of the shape representing each segment of a polypeptide tied by hydrogen bonds, van der Walls forces, electrostatic interaction and hydrophobic effects. It is also formed around a group of amino acids considered as the ground state, and extended to include adjacent amino acids till the blocking amino acids are reached and the whole protein chain along the polypeptide adopted its preferred secondary structure. However, such mechanism has not yet been understood at the satisfactory level. For instance, the studies based on statistical analysis of identifying the probabilities of locating amino acids in each secondary structure are still at the level of less than 75\% accuracy. Moreover, the main mechanism responsible for a structured folding pathway have also not yet been identified at all. Although, it is believed that such protein misfolding has been identified as the main cause of several diseases like cancers and so on \cite{dobson}.

On the other hand, some previous studies have shown that the nonlinear
excitations could play an important role in conformational dynamics of protein
backbone by decreasing the effective binding rigidity of a biopolymer chain
leading to a buckling instability of the chain \cite{mingaleev}. The results
motivate us to develop a model describing the conformational changes of protein
based on the $\phi^4$ theory \cite{januar}. The model is actually inspired by
some previous models which attempt to reproduce the protein folding using
nonlinear Schr\"odinger hamiltonian with the additional tension-like force
\cite{jacob1,berloff}. Those explain the transition of a protein from a
metastable to its ground conformation induced by solitons, while the mediator of
protein transition is the Davydov solitons propagating through the protein
backbone \cite{jacob2}. It has been shown that our model could reproduce and
improve such models more naturally from first principle using lagrangian
formalism. Another known theoretical study for the conformational dynamics of
biomolecules is the so-called ab initio quantum
chemistry approach which, however requires astronomical computational power to deal with realistic biological systems \cite{garcia, onuchi}. In contrary, along with the current model there are also some attempts to describe the dynamics in term of elementary biomatter using field theory approach \cite{sulaiman} and open quantum system \cite{sulaiman2, sulaiman3}.

This paper follows the same model in \cite{januar} to describe the protein
folding dynamics. In contrast with the previous works adopting nonlinear
Schr\"odinger equation and putting the required interactions by hand, e.g.
\cite{jacob1,jacob2,berloff}, the $\phi^4$ term in the present model produces
nonlinear Klein-Gordon equation as a source of disturbance, that is the $\phi^4$
self-interaction generates the nonlinear and tension force terms naturally
\cite{sulaiman3}. In the model, the protein backbone is assumed linear for the
initial condition. Then the nonlinear bunch of the light, like a laser, passed
to the backbone. The interaction between the conformational changes and
nonlinear source leads to certain U(1) symmetry breaking. This would be the main
source of protein folding. However more than investigating its dynamics as done
in such previous works \cite{januar, berloff}, this paper deals with statistical
properties involved in the process. In particular, the focus is put on the heat
capacity in a certain volume, $C_V$, representing the energy absorption against
the temperature changes. The effect of nonlinear sources on $C_V$ is
investigated.

The paper is organized as follows. First, the model and the underlying
assumptions are briefly reviewed in detail in Sec. 2. It is then followed by the
short derivation of relevant equation of motions (EOMs) as done in
\cite{januar} and showing the refined numerical simulation of folding process
within the model. In Sec. 4 the statistical mechanics properties are
investigated in detail. Finally the paper is concluded with summary and
discussion.

\section{The Models}

Let us briefly review the model proposed in our previous work \cite{januar}. The
lagrangian density in the model is given as,
\be
\l_{tot} = \l_c(\phi)+\l_s(\psi)+\l_{int}(\phi,\psi)\; ,
\label{eq:ltot}
\ee
where, 
\bea
\l_c	& = & \frac{1}{2}\left[ \left( \pd_\mu \phi \right)^\dagger \left(
\pd^\mu \phi \right) + m_{\phi}^2 \left( \phi^\dagger \phi \right) \right] \; ,
\label{eq:Lc}\\
\l_s	& = & \frac{1}{2} \left[ \left( \pd_\mu \psi \right)^\dagger  \left(
\pd^\mu \psi \right) - \frac{1}{2} \lambda \, \left( \psi^\dagger \psi
\right)^2 \right] \; , \label{eq:Ls}\\
\l_{int} & = & \Lambda \, \left( \phi^\dagger \phi \right) \left( \psi^\dagger
\psi \right) \; ,
\label{eq:Li}
\eea
representing the conformational changes of a protein backbone and the nonlinear
source injected to the backbone, while the last one is the interaction term 
between both. 

From the lagrangian, the total potential working in the system can be written
as, 
\be
 V(\psi,\phi) = -\frac{\lambda}{4} \, \left( \psi^\dagger \psi \right)^2 
	  + \Lambda \, \left( \phi^\dagger \phi \right) \left(\psi^\dagger
\psi \right) \; . 
  \label{eq:v}
\ee
Imposing a local U(1) symmetry to the total lagrangian and considering its
minima lead to the vacuum expectation value (VEV), 
\be
  \langle \psi \rangle = \sqrt{\frac{2\Lambda}{\lambda}} \langle \phi
\rangle \; .
  \label{eq:vev}
\ee
This non-zero VEV then yields the so-called spontaneous symmetry breaking. On
the other hand, substituting $\langle \psi \rangle$ into Eq. (\ref{eq:v})
induces the 'tension force' which plays an important role to enable folded
pathways appear naturally.

The symmetry breaking at the same time shifts the $\phi$ mass as follow,
\be
  m_{\phi}^2 \rightarrow \overline{m}_{\phi}^2 \equiv m_{\phi}^2 - \frac{2\Lambda^2}{\lambda} \langle \phi \rangle^2 \; .
\ee
Roughly, $\langle \phi \rangle$ and $m_\phi$ are at the same order. Then one can
obtain a constraint for the couplings as follow,
\be
  1 - \frac{2\Lambda^2}{\lambda}  > 0 \; \; 
  \mathrm{or} \; \; 2 \Lambda^2 < \lambda \; ,
\label{eq:constrain}
\ee
to guarantee the positive masses.

Now we are ready to move further on investigating the dynamics and statistical
properties of protein folding within the model.

\section{Dynamics of EOMs}

Having the total lagrangian in Eq. (\ref{eq:ltot}) at hand, one can derive
immediately respective EOMs using Euler-Lagrange equation in term of $\psi$ and
$\phi$ \cite{januar},
\bea
  \left( \frac{\partial^2}{\partial x^2} - \frac{1}{c^2}
\frac{\partial^2}{\partial t^2} - \frac{1}{\hbar^2} m_{\phi}^2 c^2 - 2
\Lambda \, \psi^2 \right) \phi & = & 0 \; , 
  \label{eq:eomc}\\
  \left( \frac{\partial^2}{\partial x^2} - \frac{1}{c^2}
\frac{\partial^2}{\partial t^2} - 2 \Lambda \, \phi^2 + \lambda \, \psi^2 
\right) \psi & = & 0 \; .
  \label{eq:eoms}
\eea
Note that from now the natural unit is restored to make the light velocity
($c$) and $\hbar$ appear explicitly in the equations.

Since the EOMs  under consideration are coupled nonlinear partial differential
equations, then one should in principle solve them numerically. The numerical
solution are done using the forward finite difference method \cite{fink}.
Both coupled EOMs in Eqs. (\ref{eq:eomc}) and (\ref{eq:eoms}) are rewritten in
explicit discrete forms as follows,
\bea
u_{i,j+1} & = & 2u_{i,j}-u_{i,j-1}+c^2\epsilon^2
\left( \frac{u_{i+1,j}-2u_{i,j}+u_{i-1,j}}{\delta^2} - 2 \Lambda
w_{i,j}^2u_{i,j} + \lambda u_{i,j}^3 \right),
\label{eq:u}\\
w_{i,j+1} & = & 2w_{i,j}-w_{i,j-1}+c^2\epsilon^2 
\left( \frac{w_{i+1,j}-2w_{i,j}+w_{i-1,j}}{ 
\delta^2} - 2\Lambda u_{i,j}^2w_{i,j}
\right.\nonumber\\
&&\left. -\frac{c^2}{\hbar^2}m_{\phi_c}^2w_{i,j} \right) \; ,
\label{eq:w}
\eea
for $i = 2, 3, \cdots, N-1$ and $j = 2, 3, \cdots, M-1$. In finite difference scheme, it is more convenient to replace $\psi$ and $\phi$ with {\it u} and {\it w} respectively. 
The following boundary conditions for both fields must be deployed,
\be
\begin{array}{lcl}
 \psi(0,t) = \psi(L,t)=0\;\;\mathrm{and}\;\;\phi(0,t)=\phi(L,
t)=0 & \mathrm{for} & 0\le t \le b \; , \\
\psi(x,0)=f(x)\;\;\mathrm{and} \;\;\phi(x,
0)=p(x) & \mathrm{for} & 0\le x\le L \; , \\
\displaystyle \frac{\partial\psi(x,0)}{\partial t} =
g(x)\;\;\mathrm{and}\;\;\frac{\partial\phi(x,0)}{\partial
t}=q(x) & \mathrm{for} & 0 < x < L \; ,
\end{array}
\label{eq:bc}
\ee
with $f(x)$, $p(x)$, $g(x)$ and $q(x)$ are newly introduced auxiliary functions.
The discretized value between these boundary conditions consists of $(N-1)
\times (M-1)$ rectangles with side length $\Delta x=\delta$ and $\Delta
t=\epsilon$, where the side lengths must be very small to reduce truncation
error. 

In order to calculate Eqs. (\ref{eq:u}) and (\ref{eq:w}) across the whole
region, two lowest initial values must be given. On the other hand, the value at
$t_1$ is fixed by the boundary conditions in Eq. (\ref{eq:bc}). The second order
of Taylor expansion can also be used to determine the values in the second row.
Therefore, the values at $\it t_2$ are determined by, 
\bea
u_{i,2} & = & f_i-\epsilon g_i+\frac{c^2\epsilon^2}{2} 
\left( \frac{f_{i+1}-2f_i+f_{i-1}}{\delta^2} - 2 \Lambda p_i^2 f_i + \lambda
f_i^3 \right) \; , 
\label{eq:u2}\\
w_{i,2} & = & p_i-\epsilon q_i+\frac{c^2\epsilon^2}{2} 
\left( \frac{p_{i+1}-2p_i+p_{i-1}}{\delta^2} - 2 \Lambda
f_i^2p_i-\frac{c^2}{\hbar^2}m_{\phi_c}^2p_i \right) \; ,
\label{eq:w2}
\eea
for $i = 2, 3, \cdots, N-1$. Initially, let us assume that the nonlinear source
has a particular form of $f(x)=2 \mathrm{sech}(2x) \, \mathrm{e}^{i2x}$ and
$g(x) = 1$ to generate the $\alpha$-helix, while $q(x) = 0$ for the sake
of simplicity. Then, one can obtain the next lowest initial values in this case
using Eqs. (\ref{eq:u2}) and (\ref{eq:w2}). The subsequent values are generated
by substituting the
preceding values into Eqs. (\ref{eq:u}) and (\ref{eq:w}). The higher order
values can be obtained using iterative procedure.

\begin{figure}[t]
        \centering 
	\includegraphics[angle=90,width=140mm,height=\textheight]{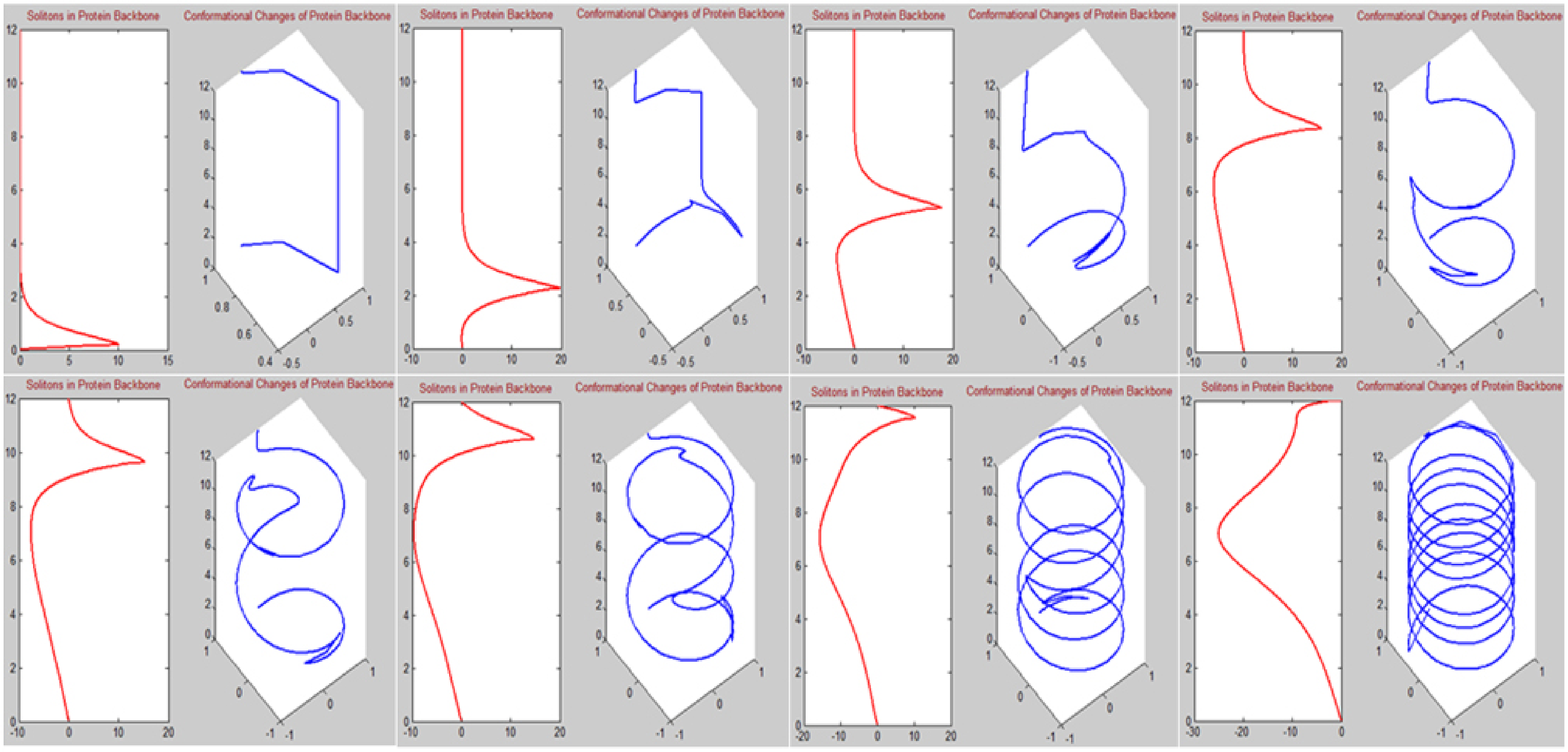}

        \caption{The soliton propagations and conformational changes on
the protein backbone inducing protein folding. The vertical axis in soliton
evolution denotes time in second, while the horizontal axis denotes its
amplitude. The conformational changes are on the $(x,y,z)$ plane.}
        \label{fig:biopolimer}
\end{figure}

In this paper, the simulation is done using the following values for the
relevant parameters (in natural units) : $m = 0.08$ eV, $L = 2.364$
nm, $\Lambda = 0.0028$, $\lambda = 0.0003$. It should be emphasized that these
values satisfy the constrain in Eq. (\ref{eq:constrain}). The result is given
in Fig. \ref{fig:biopolimer}. This result is also a revised version of the
previous one reported in \cite{januar} which contains some technical errors,
although the conclusion remains the same.

The left figure in each box describes the propagation of nonlinear sources in
protein backbone, while the right one shows how the protein is folded. As can be
seen in the figure, the protein backbone is initially linear before the
nonlinear source injection. As the soliton started propagating over the
backbone, the conformational changes appear. It should be remarked that the
result is obtained up to the second order accuracy in Taylor expansion. In order
to guarantee that the numerical solutions do not contain large amount of
truncation errors, the step sizes $\delta$ and $\epsilon$ are kept small enough.
Nevertheless, this should be good approximation to describe visually the
mechanism of protein folding. 

\section{Statistical mechanics}

Now let us discuss the main part of this paper. The statistical properties of a
system with a particular lagrangian can be investigated through its
partition function. It should be emphasized that the statistical observables
hold only on an equilibrium which is fortunately guaranteed in the present case
since the lagrangian under consideration is just the well known Klein-Gordon
scalar lagrangian.

The statistical observables can be conveniently calculated from the generating
functional by the perturbation method \cite{mackenzie}. The generating
functional for scalar fields is written as,
\be
Z = \int \mathcal{D}\phi\mathcal{D}\psi \, \exp\left\{i\int d^2x 
\l_{tot}(\phi,\psi)\right\} \; .
\ee
The partition function can further be obtained from the generating functional
by implementing a Wick rotation of the real axis \cite{kapusta}, \ie by
defining the imaginary time $i t = \tau$. Considering the finite time, the
integral is performed between the range of $-\beta/2 \sim \beta/2$ with a
periodicity condition of the field, that is $\phi ( 0,-\frac{\beta}{2} ) = \phi
( L,\beta/2 )$. Here, $L$ is a fixed boundary of one dimensional space of
protein backbone, while $\beta = 1/T$ with $T$ is the absolute temperature in
Kelvin. This specifically leads to the finite temperature case in Euclidean
coordinates,
\be
Z = \int \mathcal{D}\phi\mathcal{D}\psi \, \exp\left\{ 
\int_0^{\beta} \int_0^L d\tau dx \, \l_{tot}(\phi,\psi)\right\}\;.
\label{eq:generate}
\ee

Following standard prescription in field theory, let us consider the vacuum
transition amplitude in the presence of sources $J(x)$'s. In this approach, the
interactions can be represented by linear forms of the sources in term of the
free particle lagrangian $\l_{0}$, 
\be
Z_{0}[J_{\psi}(x),J_{\phi}(x)] = \int \mathcal{D} \phi \mathcal{D} \psi \,
\exp\left\{\int d^2x\left[ \l_{0}(\phi,\psi) + J_{\phi}(x) \phi(x) +
 J_{\psi}(x) \psi(x) \right] \right\} \; ,
\label{eq:Z0}
\ee
where $\l_{0}(\phi,\psi)= \frac{1}{2}\pd_\mu\phi\pd^\mu\phi +
\frac{1}{2}m_{\phi}^2 \phi^2+\frac{1}{2}\pd_\mu \psi\pd^\mu \psi$ and $\int
d^2x = \int_0^{\beta} \int_0^L d\tau dx$. Thereafter, the desired interactions
are derived by taking its derivatives up to certain power with respect to $J$'s
at zero points.

For instance, the $\psi^4$ term in Eq. (\ref{eq:Ls}) can be derived through the
4th derivative of $Z_0$ in Eq. ({eq:Z0}) with respect to $J_\psi$ at
$J_{\phi,\psi} = 0$,
\be
\left.\frac{\delta^4 Z_{0}}{\delta J_{\psi}^4(x)}\right|_{J_{\phi}=0,J_{\psi}=0}
=Z_0\psi^{4}(x)\;.
\label{eq:4FD}
\ee
One can perform the same procedure to obtain another interaction terms,
\be
\left.\frac{\delta^4 Z_{0}}{\delta J_{\phi}^2\delta
J_{\psi}^2}\right|_{J_{\phi}=0,J_{\psi}=0} =Z_0\phi^{2}\psi^{2} \; ,
\ee
and so forth. This means one can represent the interactions terms in term of
differential functional operators which then simplify the complete generating
functional to be,
\bea
Z&=&\exp\left\{\int d^2x\left( -\frac{\lambda}{4}\frac{\delta^4}{\delta
J_{\psi}^4} + \left.\Lambda\frac{\delta^4}{\delta J_{\phi}^2\delta
J_{\psi}^2}\right)\right|_{J_{\phi}=0,J_{\psi}=0}\right\}Z_{0}[J_{\phi},J_{\psi}
]\;.
\label{eq:FDF}
\eea

\subsection{Partition function}

The integral in Eq. (\ref{eq:generate}) can be evaluated analytically using the
Gaussian integral. This can be accomplished by rewriting it in term of Gaussian
integral using the Fourier representation of Green's function, that
is \cite{ryder}, 
\be
 (\partial_\mu\phi) (\partial^\mu\phi) = - \phi\square\phi, \;\;\mathrm{and}
\;\; (\partial_\mu\psi) (\partial^\mu\psi) = - \psi\square\psi \; ,
\label{ga}
\ee
with the D'Alembertian stands for, 
\be
\square\equiv-\frac{\pd^2}{\pd t^2}-\frac{\pd^2}{\pd x^2} \;.
\ee
Substituting this result into Eq. (\ref{eq:Z0}) yields,
\bea
Z_0 & = & \int \mathcal{D}\phi \mathcal{D}\psi \, \exp \left \{\int d^2x
\left[ - \frac{1}{2} \phi \left( \square + m^2_{\phi} \right) \phi - 
\frac{1}{2} \psi \square \psi + J_{\phi} \phi + J_{\psi} \psi \right] 
\right\} \; .
\label{eq:Z2}
\eea

Throughout the paper, for the sake of simplicity the fields are composed by its
mean values corresponding to its classical trajectories and the quantum 
fluctuations around the mean value. Therefore, the fields can be expanded
as \cite{feynman},
\bea
\phi & \equiv & \bar\phi(x) \; , \\
\psi & \equiv & \bar\psi(x)+\psi^\prime(x) \; ,
\eea
where $\bar\phi$ and $\bar\psi$ are the mean fields of the classical path while
$\psi'$ is the dispersion of solutions. The variation of conformational field
$\phi'$ is considered to be much less significant in the system. This is
motivated by a fact that the protein is a classical matter with an infinitesimal
dispersion relative to its mean value, \ie its quantum aspect is negligible 
($\phi'=0$).  The expression in Eq. (\ref{eq:Z2}) becomes, 
\bea
Z_0 & = & \int \mathcal{D}\phi \mathcal{D}\psi \, \exp \left 
\{-\int d^2x \left[ \frac{1}{2} 
\bar\phi \left( \square + m_{\phi}^2 \right) \bar\phi - J_\phi \bar\phi + 
\frac{1}{2} \bar\psi \square \bar\psi \right. \right. \nonumber\\
 & & \left. \left.
 + \frac{1}{2} \bar\psi \square
\psi' + \frac{1}{2} \psi' \square \bar\psi + 
\frac{1}{2} \psi' \square \psi' - J_\psi \bar\psi - J_\psi \psi' \right]
\right\} \; .
\eea

Using similar argument to obtain Eq. (\ref{ga}), one can apply the relation
$\int \bar\psi\square\psi' d^2x = \int \psi'\square\bar\psi d^2x$ to get,
\bea
Z_0&=&\int \mathcal{D}\phi \mathcal{D}\psi\exp\left\{-\int d^2x\left(\frac{1}{2}
\bar\phi\left(\square+m_{\phi}^2\right)\bar\phi-J_\phi \bar\phi + \frac{1}{2}
\bar\psi\square\bar\psi \right.\right.\nonumber\\
&&\left.\left.+\;\psi'\square\bar\psi+\frac{1}{2}
\psi'\square\psi'-J_\psi \bar\psi-J_\psi \psi' \right] \right\} \; .
\label{eq:Z3}
\eea
The classical path must satisfy the classical EOMs that are obtained from
the lagrangian, 
\be
\square\bar\psi(x) = J_\psi(x) \; \; \; \; \; \mathrm{and} \; \; \; \; \; 
\left( \square + m_{\phi}^2 \right) \bar\phi(x) = J_\phi(x) \; ,
\label{eq:phi0}
\ee
with the solutions,
\be
\bar\psi(x) = \int \Delta_{\psi}(x-y)J_{\psi}(y)d^2y 
\; \; \; \; \; \mathrm{and} \; \; \; \; \; 
\bar\phi(x) = \int \Delta_{\phi}(x-y)J_{\phi}(y)d^2y \; ,
\label{eq:phi01}
\ee
where $\triangle(x-y)$ is the Feynman propagator. Substituting Eqs.
(\ref{eq:phi0}) and (\ref{eq:phi01}) into Eq. (\ref{eq:Z3}) yields,
\bea
Z_0&=&\exp\left\{\frac{1}{2}\int
d^2xd^2y\left[J_{\phi}(x)\Delta_{\phi}(x-y)J_{\phi}(y)+J_{\psi}(x)\Delta_{\psi}
(x-y)J_{\psi}(y)\right]\right\}\nonumber\\
&&\times\;\int\mathcal{D}\psi'\exp\left\{-\int d^2x\frac{1}{2}\psi'\square\psi'\right\}\;.
\label{eq:Z0J1}
\eea
Under this approximation, only $\psi'$ remains in the path integral and the
result is just a number, namely $N$.

Now the remaining task is calculating the  transition amplitude by
considering the Taylor expansion of Eq. (\ref{eq:Z0J1}),
\bea
Z_0&=&N\left\{1+\frac{1}{2}\int d^2xd^2y \left[J_{\phi}(x)\Delta_{\phi}(x-y)J_{\phi}(y)+J_{\psi}(x)\Delta_{\psi}(x-y)J_{\psi}(y)\right]\right.\nonumber\\ &&\left.
 + \frac{1}{2!}\left(\frac{1}{2}\right)^2  \right. \nonumber \\
 && \left. \times \left(\int d^2xd^2y\left[
J_{\phi}(x)\triangle_{\phi}(x-y)J_{\phi}(y)+J_{\psi}(x)\triangle_{\psi}(x-y)J_{
\psi}(y)\right]\right)^2 \right.\nonumber\\
&&\left.+ \;\cdots\right\}\;.
\eea
Considering the higher order derivatives to retrieve the interaction terms as
discussed before, the survived terms are,
\bea
Z_0 & \approx & N \frac{1}{2!} \left(\frac{1}{2}\right)^2 \nonumber \\
 && \times \left(\int
 d^2xd^2y\left[
J_{\phi}(x)\Delta_{\phi}(x-y)J_{\phi}(y)+J_{\psi}(x)\Delta_{\psi}(x-y)J_{\psi}
(y)\right]\right)^2 \; .
\label{eq:Z0J}
\eea
This result is substituted into Eq. (\ref{eq:FDF}) to get,
\be
Z =  N \, \ex^{\zeta} \, \frac{\kappa^2}{8} \;, 
\ee
where,
\bea
\zeta & = &\int d^2x\left( -\frac{\lambda}{4}\frac{\delta^4}{\delta
J_{\psi}^4(x)} + \Lambda\frac{\delta^4}{\delta J_{\phi}^2(x)\delta 
J_{\psi}^2(x)}\right)\;,\\  
\kappa&=&\int d^2x_1d^2x_2\left[
J_{\phi}(x_1)\Delta_{\phi}(x_1-x_2)J_{\phi}(x_2)+J_{\psi}(x_1)\Delta_{\psi}
(x_1-x_2)J_{\psi}(x_2)\right]\;,\\
\label{eq:an}
\eea

With these notations, one can evaluate the survived term in $\ex^{\zeta}
\kappa^2$, 
\bea
\zeta \kappa^2&=&\int d^2x\left\{-\frac{\lambda}{4}\frac{\delta^4\kappa^2}{\delta
J_{\psi}^4(x)} + \Lambda\frac{\delta^4\kappa^2}{\delta J_{\phi}^2(x)\delta
J_{\psi}^2(x)}\right\}\nonumber\\
&=&\int
d^2x\left\{ -6\lambda\Delta_{\psi}^2(0) + 8 \Lambda \Delta_{\phi}(0)
\Delta_{\psi} (0)\right\} \; .
\eea
since,
\bea
\dot\kappa_{\psi}&=&\frac{\delta \kappa}{\delta J_{\psi}(x)}=\int
d^2x_2\Delta_{\psi}(x-x_2)J_{\psi}(x_2)+\int
d^4x_1J_{\psi}(x_1)\Delta_{\psi}(x_1-x)\;,\\
\dot\kappa_{\phi}&=&\frac{\delta \kappa}{\delta J_{\phi}(x)}=\int
d^2x_2\Delta_{\phi}(x-x_2)J_{\phi}(x_2)+\int
d^2x_1J_{\phi}(x_1)\Delta_{\phi}(x_1-x)\;,\\
\ddot\kappa_{\psi}&=&\frac{\delta^2 \kappa}{\delta
J_{\psi}^2(x)}=2\Delta_{\psi}(0)\;,\\
\ddot\kappa_{\phi}&=&\frac{\delta^2 \kappa}{\delta
J_{\phi}^2(x)}=2\Delta_{\phi}(0)\;.
\eea
Finally, this leads to,
\bea
Z & = & N \exp \left\{ \int_0^{\beta} d\tau \int_0^L dx 
\left[ -\frac{3}{4}\lambda\Delta_{\psi}^2(0) +  
\Lambda\Delta_{\phi}(0)\Delta_{\psi}(0)\right]\right\} \; .
\eea
This is the master equation to investigate the statistical properties in the
next subsection.

\subsection{Statistical observables}

Let us consider the specific heat of the system in a constant volume, $C_V$,
that is a particular interest from experimental point of view. The specific heat
can be derived directly from the partition function using the relation, 
\be
C_V = \beta^2 \, \left( \frac{\partial^2 \ln Z}{\partial \beta^2} \right)_V \;
. 
\ee
In the present case, it is found to be,
\bea
C_V & = & \beta^2 \, \frac{\partial^2}{\partial \beta^2} \left(\ln{N} - 
 \beta L \left[ \frac{3}{4}\lambda\Delta_{\psi}^2(0)
- \Lambda\Delta_{\phi}(0)\Delta_{\psi}(0)\right]\right) \; ,
\label{eq:omega}
\eea
after performing the integration over $\tau$ and $x$ respectively, while the
overall factor $N$ has been obtained as \cite{mackenzie},
\be
  N = \frac{1}{4 \pi \sinh({k \beta}/2)}.
\ee

Next, one must find out the form of $\Delta(0)$. This can be achieved by solving
the the Green function. Since the Feynman propagator $\Delta(x)$ obeys,
\be
\begin{array}{lcl}
\displaystyle \left(-\frac{\partial^2}{\partial
\tau^2}-\frac{\partial^2}{\partial x^2}\right) \Delta_{\psi}(x,\tau) & =
&\delta(x)\delta(\tau) \;,\\
\displaystyle \left(-\frac{\partial^2}{\partial
\tau^2}-\frac{\partial^2}{\partial x^2}+m^2_{\phi}\right) \Delta_{\phi}(x,\tau)
& = & \delta(x)\delta(\tau) \; ,
\end{array}
\label{eq:green}
\ee
and taking a particular form of Green functions \cite{bellac},
\be
\begin{array}{lcl}
\Delta_{\psi}(x,\tau)&=&\displaystyle \int\frac{dk}{2\pi}e^{ikx}\Delta_{\psi}
(\tau)\;,\\
\Delta_{\phi}(x,\tau)&=&\displaystyle \int\frac{dq}{2\pi}e^{iqx}\Delta_{\phi}(\tau)\; ,
\end{array}
\label{eq:delta}
\ee
the imaginary-time propagators $\Delta(\tau)$ should satisfy the following
differential equations,
\be
\begin{array}{lcl}
\displaystyle \left(-\frac{\partial^2}{\partial \tau^2}+k^2\right)
\Delta_{\psi}(\tau)&=&\delta(\tau)\; ,\\
\displaystyle \left(-\frac{\partial^2}{\partial \tau^2}+q^2+m^2_{\phi}\right)
\Delta_{\phi}(\tau)&=&\delta(\tau) \; .
\end{array}
\label{eq:gt}
\ee
Imposing the Dirichlet periodic boundary conditions,
\be
\Delta \left(-\frac{\beta}{2}\right) = \Delta\left(\frac{\beta}{2}\right)
 \; \; \textmd{and} \; \; 
\dot\Delta \left(-\frac{\beta}{2} \right) = \dot\Delta \left(\frac{\beta}{2}
\right) \; , 
\label{eq:bgt}
\ee
and using Eq. (\ref{eq:gt}), Eq. (\ref{eq:delta}) becomes \cite{rattazzi},
\bea
\Delta_{\psi}(\tau)&=&\frac{\cosh\left(k\left(\beta/2 - |\tau|
\right)\right)}{2 k \sinh\left( {k\beta}/2 \right)}\;,\\
\Delta_{\phi}(\tau)&=&\frac{\cosh\left(\sqrt{q^2+m^2}\left(\beta/2 
-|\tau|\right)\right)}{2\sqrt{q^2+m^2}
\sinh\left({\beta\sqrt{q^2+m^2}}/2\right)}\;.
\eea
Therefore the Fourier representation of Green functions can be written as
\bea
\Delta_{\psi}(x,\tau)&=&\int{\frac{dk}{2\pi}\frac{
\ex^{ikx} \, \cosh\left(k\left(\beta/2 - |\tau| \right)\right)}{2k \, 
\sinh\left({k\beta}/2\right)}}\;,
\label{eq:dpsi}\\
\Delta_{\phi}(x,\tau)&=&\int{\frac{dq}{2\pi}\frac{\ex^{iqx}\cosh\left(\sqrt{
q^2+m^2}\left( \beta/2 - |\tau| \right)\right)}{2 \sqrt{q^2+m^2} \, 
\sinh\left({\beta\sqrt{q^2+m^2}}/2 \right)}}\;.
\label{eq:dphi}
\eea

\begin{figure}[t]
        \centering 
	\includegraphics[width=0.75 \textwidth]{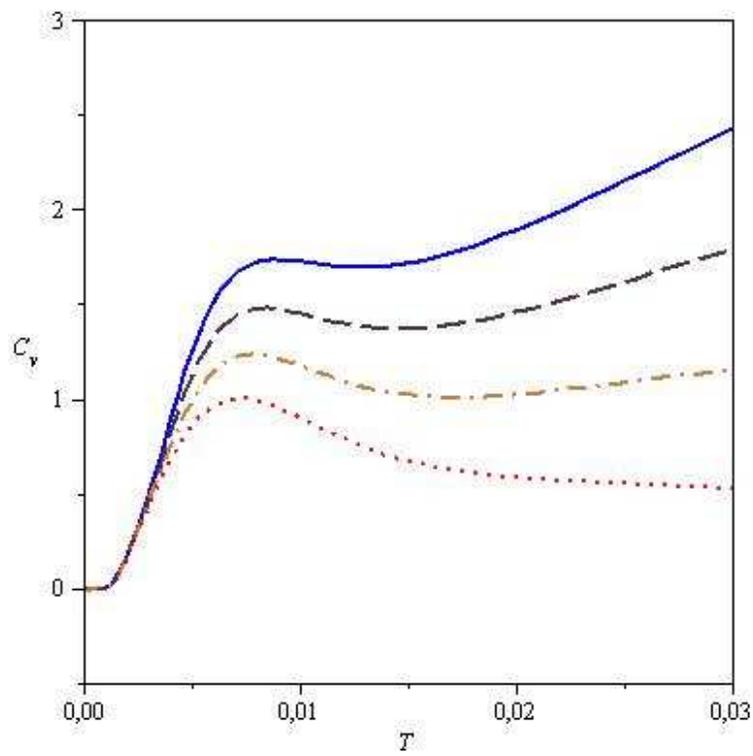}
        \caption{$C_V$ as a function of temperature for various values of
$\lambda = 0.0003$ (solid-line), 0.0012 (dashed-line), 0.0021
(dashed-dotted-line), 0.0030 (dotted-line) with $\Lambda = 0.00283$.}
        \label{fig:cv1}
\end{figure}

\begin{figure}[t]
        \centering 
	\includegraphics[width=0.75 \textwidth]{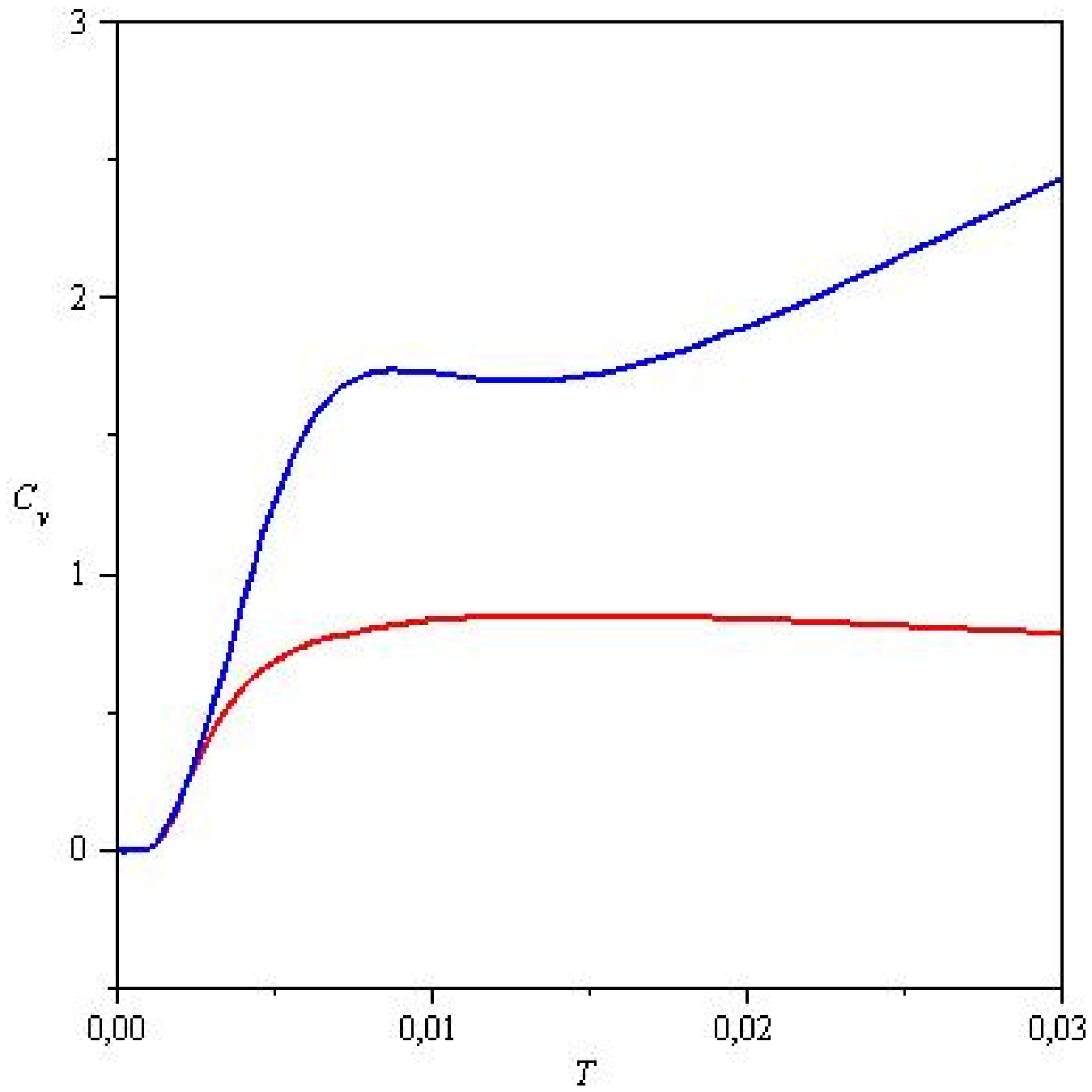}
        \caption{$C_V$ as a function of temperature for $\Lambda = 0.00283$
(blue-line) and 0 (red line) with $\lambda = 0.0003$.}
        \label{fig:cv2}
\end{figure}

As can be seen, the propagator $\Delta_{\phi}(x,\tau)$ in Eq. (\ref{eq:dphi})
contains a factor of $\sqrt{q^2+m_{\phi}^2}$ which makes the integral cannot be
performed analytically. In this paper, it is done numerically.

Performing the integrals in Eqs. (\ref{eq:dpsi}) and (\ref{eq:dphi}), and
substituting the results into Eq. (\ref{eq:omega}), one then obtains
the numerical results as shown in Figs. \ref{fig:cv1} and \ref{fig:cv2} for
various values of $\lambda$ and $\Lambda$, while $k = 0.01$. The values of
another variables are the same as in Fig. \ref{fig:biopolimer}.

\section{Conclusion}

An extension of the phenomenological model describing the conformational
dynamics of proteins has been briefly reintroduced. The model based on the
matter interactions among the conformational field and the nonlinear sources
represented as the scalar bosonic fields $\phi$ and $\psi$. As already shown
in our previous work \cite{januar}, the nonlinear and tension force terms
appear naturally from the scalar lagrangian with $\psi^4$ self-interaction.
Moreover, such forces are realized as a consequence of symmetry breaking.

In the present paper, the numerical simulation of protein folding dynamics has
been refined to revise some technical errors in the previously reported result
\cite{januar}. However, the figure has only changed slightly, while the
conclusion remains the same.

Moreover, in the present paper the statistical properties of protein folding
within the model are studied in detail. In particular, the specific heat, $C_V$,
has been calculated analytically using statistical mechanics and path integral
method. The evolution of $C_V$ in term of temperature has been shown for various
levels of nonlinearity and interaction with nonlinear source represented by
$\lambda$ and $\Lambda$. It is found that both of them contribute in an
opposite way, and could completely cancel each other at certain values as Eq.
(\ref{eq:vev}) is fulfilled. This occurs when the symmetry is maximally
broken. This also means that increasing energy absorption prefers high level
of nonlinearity of sources and at the same time weak interaction between
the sources and protein backbone.

\section*{Acknowledgments}
AS thanks the Group for Theoretical and Computational Physics LIPI for warm
hospitality during the work. This work is funded by the Indonesia Ministry of
Research and Technology and the Riset Kompetitif LIPI in fiscal year 2011 under
Contract no.  11.04/SK/KPPI/II/2011.

\bibliographystyle{elsarticle-num}
\bibliography{protein}

\end{document}